# Electromagnetically Induced Transparency using six level atoms doped in the crystalline medium

Hassan Kaatuzian; Sina Mehrabadi[*]; Ahmad Ajdarzadeh Oskouei
Electrical Engineering Department, Amirkabir University of Technology, (15914), Tehran, Iran
Emails: hsnkato@aut.ac.ir, sina_mehra@yahoo.com, ahmad_ajdarzadeh@yahoo.com

***Abstract:*** *Using the density matrix theory of interaction between light and matter, and relevant parameters of the relaxation rates for a six-level model, we have shown theoretically the possibility of realizing Electromagnetically Induced Transparency (EIT) and Slow Light effects in $Pr^{+3}$ doped $Y_2SiO_5$ crystal $(Pr^{+3}:YSO)$. In addition, we have presented a simplified method to analyse EIT effect in such a six level atomic system. Finally, we have demonstrated results of numerical calculation and have compared them with experimental measurements reported recently.*

## 1. Introduction

Interference between alternative pathways in quantum mechanical process is a frequently studied effect in physics nowadays [1]. It has led to many interesting optical phenomena such as Electromagnetically Induced Transparency (EIT) [2-3], Lasing Without Inversion (LWI) [4-5], the enhancement of refractive index (Slow Light) [6] and the quantum optical memories [12-15]. The effect named as EIT is an induced transparency in an originally absorbing medium, experienced by a weak probe field, due to the presence of a strong coupling field. Applications of EIT in the gaseous medium are limited because of atomic motion (diffusion). In the recent years, the EIT effect has been observed experimentally in crystal [7]; however, most solid materials have relatively broad optical line widths, which limit the efficiency of EIT. In the other hand; rare-earth-doped crystals have properties similar to atomic vapours, with the advantage of no atomic movement; therefore, these materials are used for ultrahigh density optical memories and process. For a type of such materials, $[Pr^{+3}$ doped $Y_2SiO_5$ $(Pr^{+3}:YSO)]$, efficient, narrow line width EIT has been experimentally demonstrated; however, there is no clear theoretical explanation to this phenomenon. In this paper, we developed a theoretical theory and calculation using density matrices, to show how the EIT effect and Slow Light can be realized in $Pr^{+3}$ doped $Y_2SiO_5$ crystal.

## 2. Theory

To study EIT within $(Pr^{+3}:YSO)$ crystal, we use semi-classical theory of interaction between light and matter [8]. Considering a six level model as shown in Figure 1, where $\omega_P$, $\omega_C$ and $\omega_A$ are frequencies of the weak probe field, strong coupling field and auxiliary field, respectively. The complex polarizability is defined as $\vec{P} = \varepsilon_0 \chi \vec{E}$ where susceptibility is given by $\chi = \chi' + i\chi''$. Susceptibility may be obtained as [8] $\chi' + i\chi'' = \dfrac{2N\mu_{52}\rho_{52}}{\varepsilon_0 E_0}$ where $\rho_{52}$ is the density matrix element connecting level $|5\rangle$ and $|2\rangle$, $\mu_{52}$ is the dipole matrix element and N is the density of $Pr^{+3}$ ions. Noting the relations $\Delta n = n-1 = 0.5\chi'$ and $\alpha = 0.5k\chi''$ where $\Delta n$ is the refractive index coefficient and $\alpha$ is absorption coefficient of the probe field; if the density matrix element $\rho_{52}$ is known, the refractive index and absorption coefficient will be obtained. To obtain $\rho_{52}$, one should solve the time-dependent density matrix equation of motion [8]:

$$d\rho_{mn}/dt = \frac{[H_{int},\rho]}{i\hbar} \qquad (1)$$

Where $H_{int}$ is the time-independent interaction Hamiltonian (Hamiltonian matrix) in a rotating

---

[*] Corresponding Author

frame. For a system shown in Figure 1, we can write $H_{int}$ as:

$$H_{int} = -\frac{\hbar}{2}\begin{bmatrix} 0 & 0 & 0 & 0 & 0 & \Omega_A^* \\ 0 & 0 & 0 & 0 & \Omega_P^* & 0 \\ 0 & 0 & 0 & 0 & \Omega_C^* & 0 \\ 0 & 0 & 0 & 0 & 0 & 0 \\ 0 & \Omega_P & \Omega_C & 0 & 0 & 0 \\ \Omega_A & 0 & 0 & 0 & 0 & 0 \end{bmatrix} \quad (2)$$

where $\Omega_A, \Omega_P$ and $\Omega_C$ are the complex Rabi frequencies of the field coupling atomic levels of $|6\rangle - |1\rangle$, $|5\rangle - |2\rangle$ and $|5\rangle - |3\rangle$ respectively.

Assuming the perfect triple-resonance condition, the evolution equations for the density matrix element can be written as:

$$\frac{\partial}{\partial t}\rho_{11} = i/2\Omega_A^*\rho_{61} - i/2\Omega_A\rho_{66} + \quad (3)$$
$$\Gamma_{21}^{(L)}\rho_{22} + \Gamma_{31}^{(L)}\rho_{33} + \Gamma_{41}^{(L)}\rho_{44} + \Gamma_{51}^{(L)}\rho_{55} + \Gamma_{61}^{(L)}\rho_{61}$$

$$\frac{\partial}{\partial t}\rho_{22} = i/2\Omega_P^*\rho_{52} - i/2\Omega_P\rho_{25} \quad (4)$$
$$-\Gamma_{21}^{(L)}\rho_{22} + \Gamma_{32}^{(L)}\rho_{33} + \Gamma_{42}^{(L)}\rho_{44} + \Gamma_{52}^{(L)}\rho_{55} + \Gamma_{62}^{(L)}\rho_{62}$$

$$\frac{\partial}{\partial t}\rho_{33} = i/2\Omega_C^*\rho_{53} - i/2\Omega_C\rho_{35} - \Gamma_{31}^{(L)}\rho_{33} - \Gamma_{32}^{(L)}\rho_{33} + \quad (5)$$
$$\Gamma_{43}^{(L)}\rho_{44} + \Gamma_{53}^{(L)}\rho_{55} + \Gamma_{63}^{(L)}\rho_{66}$$

$$\frac{\partial}{\partial t}\rho_{55} = i/2\Omega_P\rho_{25} - i/2\Omega_P^*\rho_{52} + i/2\Omega_C\rho_{35} \quad (6)$$
$$-i/2\Omega_C^*\rho_{53} + \Gamma_{65}^{(L)}\rho_{66} - \rho_{55}(\Gamma_{51}^{(L)} + \Gamma_{52}^{(L)} + \Gamma_{53}^{(L)} + \Gamma_{54}^{(L)})$$

$$\frac{\partial}{\partial t}\rho_{66} = i/2\Omega_A\rho_{16} - i/2\Omega_A^*\rho_{61} \quad (7)$$
$$+\rho_{66}(\Gamma_{61}^{(L)} + \Gamma_{62}^{(L)} + \Gamma_{63}^{(L)} + \Gamma_{64}^{(L)} + \Gamma_{65}^{(L)})$$

$$\frac{\partial}{\partial t}\rho_{61} = i/2\Omega_A(\rho_{11} - \rho_{66}) - \gamma_{61}\rho_{61} \quad (8)$$

$$\frac{\partial}{\partial t}\rho_{53} = i/2\Omega_P\rho_{23} - i/2\Omega_C\rho_{33}$$
$$-i/2\Omega_C\rho_{53} - \gamma_{53}\rho_{53} \quad (9)$$

$$\frac{\partial}{\partial t}\rho_{52} = i/2\Omega_P\rho_{22} + i/2\Omega_C\rho_{32} \quad (10)$$
$$-i/2\Omega_P\rho_{55} - \gamma_{52}\rho_{52}$$

$$\frac{\partial}{\partial t}\rho_{32} = i/2\Omega_C^*\rho_{52} - i/2\Omega_P\rho_{35} - \gamma_{32}\rho_{32} \quad (11)$$

Where $\Gamma_{mn}^{(L)}$ is the decay rate of population between $|m\rangle$ and $|n\rangle$, $\gamma_{mn}$ is the total decay rate for $|m\rangle - |n\rangle$ containing dephasing component that is given by:

$$\gamma_{ij} = (\Gamma_i^{(L)} + \Gamma_j^{(L)} + \gamma_{ij}^{dph})/2 \quad (12)$$

Where $\gamma_{ij}^{dph}$ is the dephasing rate of the quantum coherence between $|i\rangle$ and $|j\rangle$ levels.

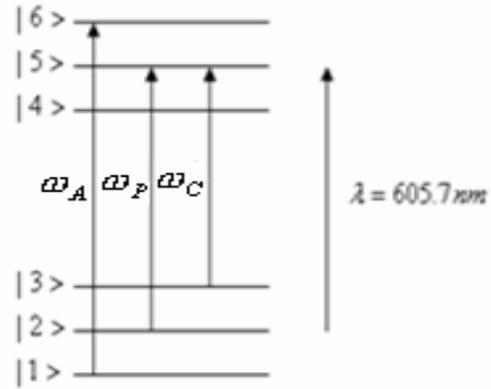

**Fig. 1** Schematics of six level atoms

## 3. Simplification

As seen earlier, the dispersion and absorption are determined by $\rho_{52}$. To calculate $\rho_{52}$, we should solve the equations (3)-(11). However, as we know the complete solution of this equation is very difficult. Therefore, it is reasonable that we simplify these equations as more as possible. Because of long life time of levels |1>, |2>, |3> (several minutes) [9] and short Lifetimes of levels |4>, |5>, |6> ($T_1 \approx 164$ μs) [9]; it is reasonable that we assume all the electronic population must be gathered into the levels |1>, |2>, |3>. In addition, assuming coupling and auxiliary fields to be strong and probe field to be weak, we may conclude that all the electronic population is concentrated in the level |2>, so we have:

$$\rho_{22} \approx 1, \quad (13)$$
$$\rho_{ii} \approx 0, \quad (i = 1,3,4,5,6) \quad (14)$$
$$\rho_{ij} \approx 0, \quad (i, j \neq 2). \quad (15)$$

Substituting these equations into equation (3-12) leads to the result that our six level atomic system is equivalent to three level $\Lambda$ system as shown in Figure 2. Therefore, our evolution equations for the elements of the density matrix is greatly simplified. For a three-level $\Lambda$ system, using slowly varying variable we have [8]:

$$\frac{\partial}{\partial t}\tilde{\rho}_{52} = -(\gamma_{52} + i\Delta)\tilde{\rho}_{52} + i/2\,\mu_{52}E_P/\hbar + i/2\Omega_C\tilde{\rho}_{32} \quad (15)$$

$$\frac{\partial}{\partial t}\tilde{\rho}_{32} = -(\gamma_{32} + i\Delta)\tilde{\rho}_{32} + i/2\Omega_C^*\tilde{\rho}_{52} \quad (16)$$

Solving these equations, we obtain:

$$\chi' = \frac{N|\mu_{52}|^2\Delta}{\varepsilon_0 Z\hbar}\left[\gamma_{32}(\gamma_{52}+\gamma_{32}) + (\Delta^2 - \gamma_{32}\gamma_{52} - \Omega_C^2/4)\right] \quad (17)$$

$$\chi'' = \frac{N|\mu_{52}|^2\Delta}{\varepsilon_0 Z\hbar}\left[\Delta^2(\gamma_{52}+\gamma_{32}) - \gamma_{32}(\Delta^2 - \gamma_{32}\gamma_{52} - \Omega_C^2/4)\right] \quad (18)$$

where Z is given by
$$Z = [(\Delta^2 - \gamma_{32}\gamma_{52} - \Omega_C^2/4)^2 + \Delta^2(\gamma_{52}+\gamma_{32})]^2 \quad (19)$$

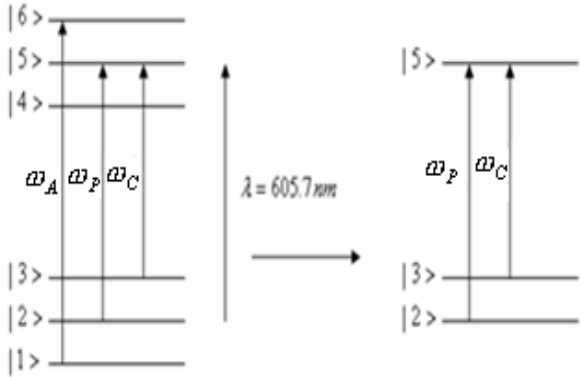

**Fig. 2** Six level atomic and its equivalent of three level $\Lambda$ atomic system.

## 4. Simulation

### 4.1 Calculation of $\gamma_{52}$ and $\gamma_{32}$

If the experimentally measured lifetime $T_1$ of the level $|i\rangle$ is known, the relaxation rate $\Gamma_i^{(L)} = 1/T_i$ may be determined. At $1.4\,°K$, the experimentally measured lifetimes of levels |1>, |2>, |3> of $Pr^{+3}$ ions in $Y_2SiO_5$ crystal are $T_1 \approx 400$ sec and the life time of level |5> is $T_1 \approx 164$ μs approximately. In addition, from ref.[9], we have $\gamma_{32}^{dph} \approx 2$ kHz and $\gamma_{52}^{dph} \approx 9$ kHz. Substituting these values in to the equation (12), we obtain $\gamma_{32} \approx 6.28\times10^3$ (rad/sec) and $\gamma_{52} \approx \gamma_{53} \approx 4.71\times10^4$ (rad/sec).

### 4.2 Numerical results

Through numerical solution of Equations (17),(18) and using the parameters given in the pervious section and setting the experimental values $N = 4.7\times10^{18}\,cm^{-3}$ [10] and $\mu_{52} \approx 10^{-33}$ [11], the $\chi'$ and $\chi''$ as a function of frequency detuning of probe field are obtained as shown in Figure 3 (In Figure 3 we have set $\Omega_C = 0$).

In the case of figure 4, we have set $\Omega_C = 1.5\,MHz$; it can be seen that a hole burning is created in the $\chi''$ profile. In this case, the width of the transparency window can be obtained approximately to be about $8\,MHz$. Because of inhomogeneous broadening, this value is larger than experimentally measured values. In addition, because the $Pr^{+3}$ atoms which occupy site 2 in crystal structure [9], the width of transparency window is limited in experiment [7].

### 4.3 Discussion on the results

The group velocity of light is given by [13]:

$$v_g = \frac{c}{n(\omega) - \omega\,dn/d\Delta} \quad (20)$$

Using equation (20) and figure 4-b; the group velocity of light can be obtained to be less than 50 m/s. This value is comparable with the values measured in experiment [7].

Considering a good agreement between the experimental results and the results obtained on our analysis, we may conclude that our simplification and analysis is consistent.

## 5. Conclusion

Using the density matrix theory of the interaction between light and matter, we have discussed the possibility to realize EIT in ($Pr^{+3}$ : YSO) with a six-level model. We have shown that the results of EIT in the case of doped atoms into the crystal in the presence of an auxiliary field, is equivalent to the results for the case of three level atomic gas. We show numerically that also in the case of doped atoms into the crystal, when the coupling field is strong enough and in the presence of auxiliary field; the absorption of the probe field is reduced drastically. In addition, when we set $\Omega_C = 1.5\,MHz$,

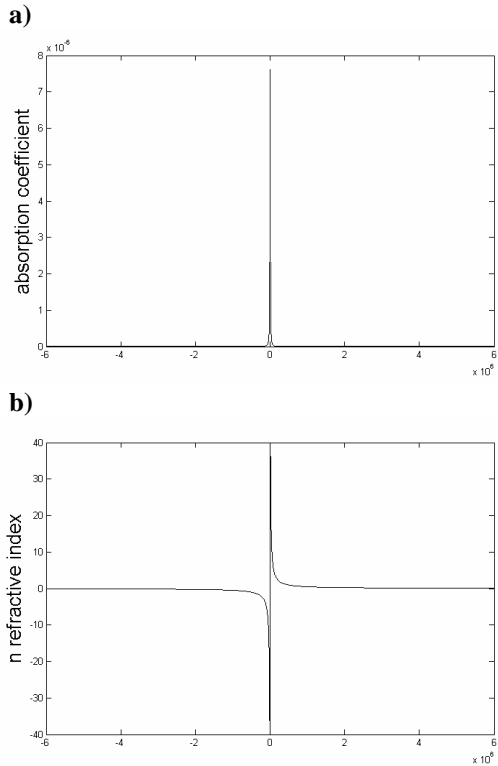

**Fig.3** a) absorption coefficient; b) refraction index (n) of the medium versus frequency detuning of the probe field $(\Delta)$, $(\Omega_C = 0)$.

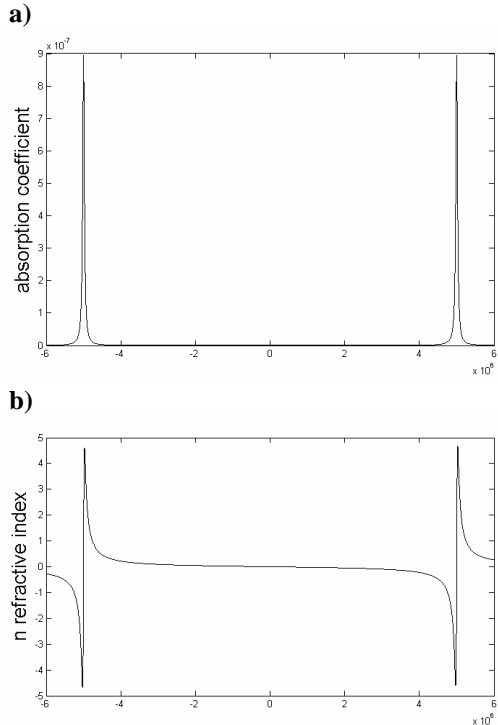

**Fig.4** a) absorption coefficient; b) refraction index (n) of the medium versus frequency detuning of the probe field $(\Delta)$, $(\Omega_C = 1.5\,MHz)$ the group velocity of the probe field is reduced to velocities less than 50 m/s; the refractive index also varies correspondingly. Considering that doped crystals (optical ionic crystals) have advantages to atomic vapour such as the characteristic of consistency and no atomic motion; EIT concept using doped atoms within the crystalline medium, can be widely used in optical storage, optical switching, phase conjugation, lasing without inversion etc.